\documentclass[conference]{IEEEtran}
\IEEEoverridecommandlockouts

\usepackage{cite}
\usepackage{amsmath,amssymb,amsfonts}
\usepackage{algorithmic}
\usepackage{graphicx}
\usepackage{textcomp}
\usepackage{xcolor}
\usepackage{url}
\usepackage{subcaption}
\usepackage{svg}
\usepackage{enumitem}
\usepackage[hidelinks]{hyperref}

\def\BibTeX{{\rm B\kern-.05em{\sc i\kern-.025em b}\kern-.08em
    T\kern-.1667em\lower.7ex\hbox{E}\kern-.125emX}}
\begin{document}

\title{Correlation and Temporal Consistency Analysis of Mono-static and Bi-static ISAC Channels}
        
\author{\IEEEauthorblockN{Saúl Fenollosa+, Narcis Cardona+, Wenfei Yang*, Jian Li*}
\IEEEauthorblockA{\textit{+Universitat Politècnica de València, Spain, *Huawei Technologies Co., Ltd., China} \\         \textit{sfenarg@upv.edu.es, ncardona@iteam.upv.es, yangwenfei4@huawei.com, calvin.li@huawei.com}}}

\maketitle

\begin{abstract}
Integrated Sensing and Communication (ISAC) is critical for efficient spectrum and hardware utilization in future wireless networks like 6G. However, existing channel models lack comprehensive characterization of ISAC-specific dynamics, particularly the relationship between mono‑static (co‑located Tx/Rx) and bi‑static (separated Tx/Rx) sensing configurations. Empirical measurements in dynamic urban microcell (UMi) environments using a 79‑GHz FMCW channel sounder help bridge this gap. Two key findings are demonstrated: (1) mono‑static and bi‑static channels exhibit consistently low instantaneous correlation due to divergent propagation geometries; (2) despite low instantaneous correlation, both channels share unified temporal consistency, evolving predictably under environmental kinematics. These insights, validated across seven real‑world scenarios with moving targets/transceivers, inform robust ISAC system design and future standardization.
\end{abstract}


\section{Introduction}

\begin{figure}[htbp]
  \centering
  \begin{subfigure}[b]{0.25\textwidth}
    \includegraphics[width=\linewidth]{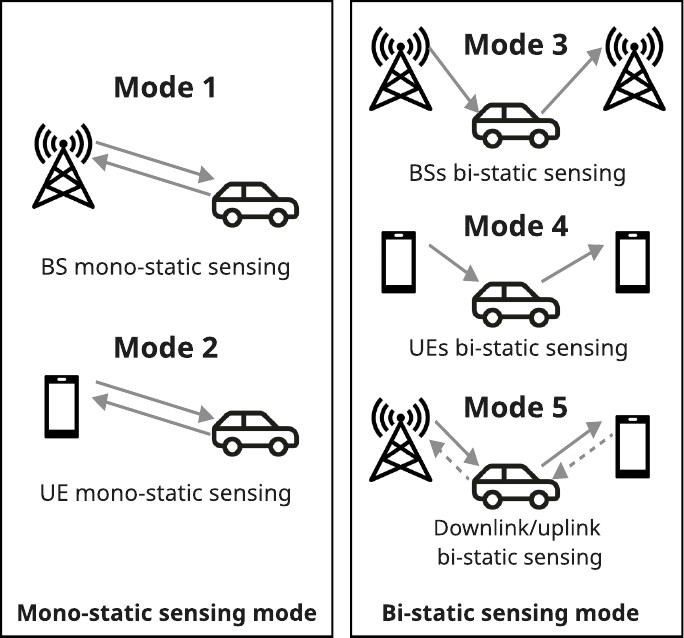}
    \caption{}
    \label{fig:isac_modes}
  \end{subfigure}\hfill
  \begin{subfigure}[b]{0.23\textwidth}
    \includegraphics[width=\linewidth]{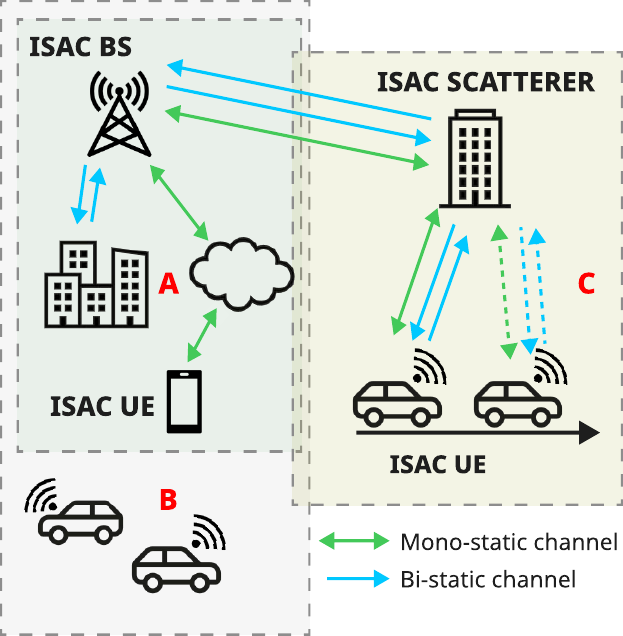}
    \caption{}
    \label{fig:isac_scenarios}
  \end{subfigure}
  \caption{Definition of ISAC Sensing Modes (a) and Channel Configurations (b)}
\end{figure}

Integrated Sensing and Communication (ISAC) represents a pivotal paradigm for future wireless networks, including 6G, by integrating sensing and wireless communication functionalities to enhance spectrum and hardware utilization \cite{Lu2024ISAC, zhang2025integratedsensingcommunicationsyears}. Effective ISAC system design fundamentally relies on precise radio channel characterization \cite{meng2024integratedsensingcommunicationmeets, wei2024integratedsensingcommunicationchannel}. However, existing channel models, such as those found in current 3GPP standards, often lack specific provisions for comprehensive ISAC channel characterization, particularly concerning distinct mono- and bi-static sensing configurations \cite{wei2024integratedsensingcommunicationchannel, heggo2025isacchannelmodelling}. Mono-static channels involve co-located transmitters and receivers, where signals undergo a round-trip propagation path to a target and back to the same point. Conversely, bi-static channels feature spatially separated transmitters and receivers, with signals traversing distinct paths from the transmitter to the target and then to a distant receiver. The inherent differences in their propagation geometries suggest that the instantaneous characteristics of these channel types may not be directly correlated \cite{ge2023integratedmonostaticbistaticmmwave, zhang2025jointbistaticpositioningmonostatic}. Building upon established classifications \cite{Wenfei2023ChanMeas}, ISAC sensing modes are primarily categorized into mono-static (where Tx and Rx are co-located) or bi-static (where Tx and Rx are separated). Mono-static Modes include BS mono-static sensing (Mode 1) and UE mono-static sensing (Mode 2). Bi-static Modes encompass BSs bi-static sensing (Mode 3), UEs bi-static sensing (Mode 4), and downlink/uplink bi-static sensing (Mode 5), as further illustrated in Figure~\ref{fig:isac_modes}. Beyond these modes, various ISAC configurations are described in Figure~\ref{fig:isac_scenarios}: Scenario A depicts static transceivers (Tx, Rx) and channel elements (e.g., targets, scatterers); Scenario B involves static Tx/Rx with dynamic channel elements; and Scenario C features dynamic Tx or Rx. This paper focuses explicitly on the dynamic environments presented in Scenarios B and C.


\vspace{-5pt}

\vspace{-5pt}
While the functional distinctions and performance trade‑offs between mono‑ and bi‑static sensing have been recognized \cite{wang2025sensingassistedchannelestimationofdm,keskin2025bridginggapdataaidedsensing}, a significant empirical gap persists in quantifying their precise relationship. Specifically, comparative analyses of their instantaneous correlation and temporal consistency in dynamic, real‑world environments remain scarce, with existing studies focusing on channel inference from sensing data or general coupling and lacking a thorough empirical examination of their synchronous temporal evolution.

This paper addresses this empirical deficiency by presenting a comprehensive measurement-based analysis of mono- and bi-static ISAC channels in diverse dynamic scenarios. Utilizing a high-frequency (E-band) Frequency Modulated Continuous Wave (FMCW) channel sounder \cite{Cardona2023Radar, Castilla2024DualISAC}, we provide novel empirical evidence that demonstrates two key findings: (1) a consistently low direct correlation between mono- and bi-static channels in terms of their instantaneous channel realizations, and (2) that despite this low instantaneous correlation, both mono- and bi-static channels exhibit a unified temporal consistency, where their evolution over time, driven by environmental kinematics, follows a similar predictable trend. These findings, derived from extensive measurement campaigns, contribute to future ISAC channel modeling standardization efforts.

\section{Channel Measurement Set-up}

\subsection{Channel Sounding System}
The channel sounding system utilized in this study is a specialized FMCW sensor-based system. This system exploits mutual interference principles to accurately measure bi-static channel characteristics, as thoroughly described in prior studies \cite{Cardona2023Radar, Castilla2024DualISAC}.

The sounding system operates in the E-band, employing a bandwidth of 4 GHz centered at 79 GHz. Each sensor unit comprises three transmit (TX) antennas and four receive (RX) antennas. Detailed specifications of the sensors are provided in Table~\ref{tab:sensor_specs}.

\begin{table}[htbp]
    \centering
    \caption{Sensor Specifications}
    \label{tab:sensor_specs}
    \begin{tabular}{|l|c|}
        \hline
        \textbf{Parameter}          & \textbf{Value}                      \\ \hline
        Frequency Range             & 76–81 GHz                           \\
        Center Frequency            & 79 GHz                              \\
        Bandwidth                   & 4 GHz                               \\
        TX / RX Antennas            & 3 / 4                                \\
        Peak Gain (per antenna)     & 10.5 dBi                            \\
        H / V Beamwidth (3 dB)      & $\pm28^\circ$ / $\pm14^\circ$ @ 78 GHz \\
        TX Power                    & 12 dBm                              \\
        Noise Figure                & 14 dB (76–77 GHz), 15 dB (77–81 GHz) \\ \hline
    \end{tabular}
\end{table}

The sensor configuration used for the measurements was selected to ensure sufficient coverage for the range demands typical of an Urban Microcell (UMi) scenario. Additionally, it was designed to maximize distance resolution by fully exploiting the 4 GHz bandwidth supported by the sensing hardware. The specific parameters of the configuration are detailed in Table~\ref{tab:sensor_configs}.

\begin{table}[htbp]
    \centering
    \caption{Measurement Configuration Parameters}
    \label{tab:sensor_configs}
    \begin{tabular}{|l|c|c|}
        \hline
        \textbf{Parameter}              & \textbf{Value} \\ \hline
        Ramp Duration ($T_\text{chirp}$)              & 112.86 µs          \\
        Chirp Slope ($S$)                           & 35.44 MHz/µs       \\
        I/Q Sampling Rate ($f_s$)                        & 18.75 MHz          \\
        Max Range (Mono/Bi-static) ($R_\text{max}$)     & 79.4 / 158.8 m     \\ \hline
    \end{tabular}
\end{table}

\subsection{Measurement scenarios}\label{sec:measurement_scenarios}

The present study is based on a comprehensive set of seven UMi measurements, all conducted under line-of-sight (LOS) conditions, and designed with dynamic configurations B and C in the vicinity of the Nexus and 6D buildings at the Universitat Politècnica de València (UPV). These measurements aim to characterize Mono-static Modes 1 and 2, as well as Bi-static Mode 5, through the BS-UE link, in dynamic channels. Although the ISAC target in all experiments was a car, its model was not consistent across measurements, as detailed in subsequent sections.

\vspace{3pt}
\subsubsection{Scenarios B}
\begin{figure}[htbp]
  \centering
  \begin{subfigure}[b]{0.2506\textwidth}
    \includegraphics[width=\linewidth]{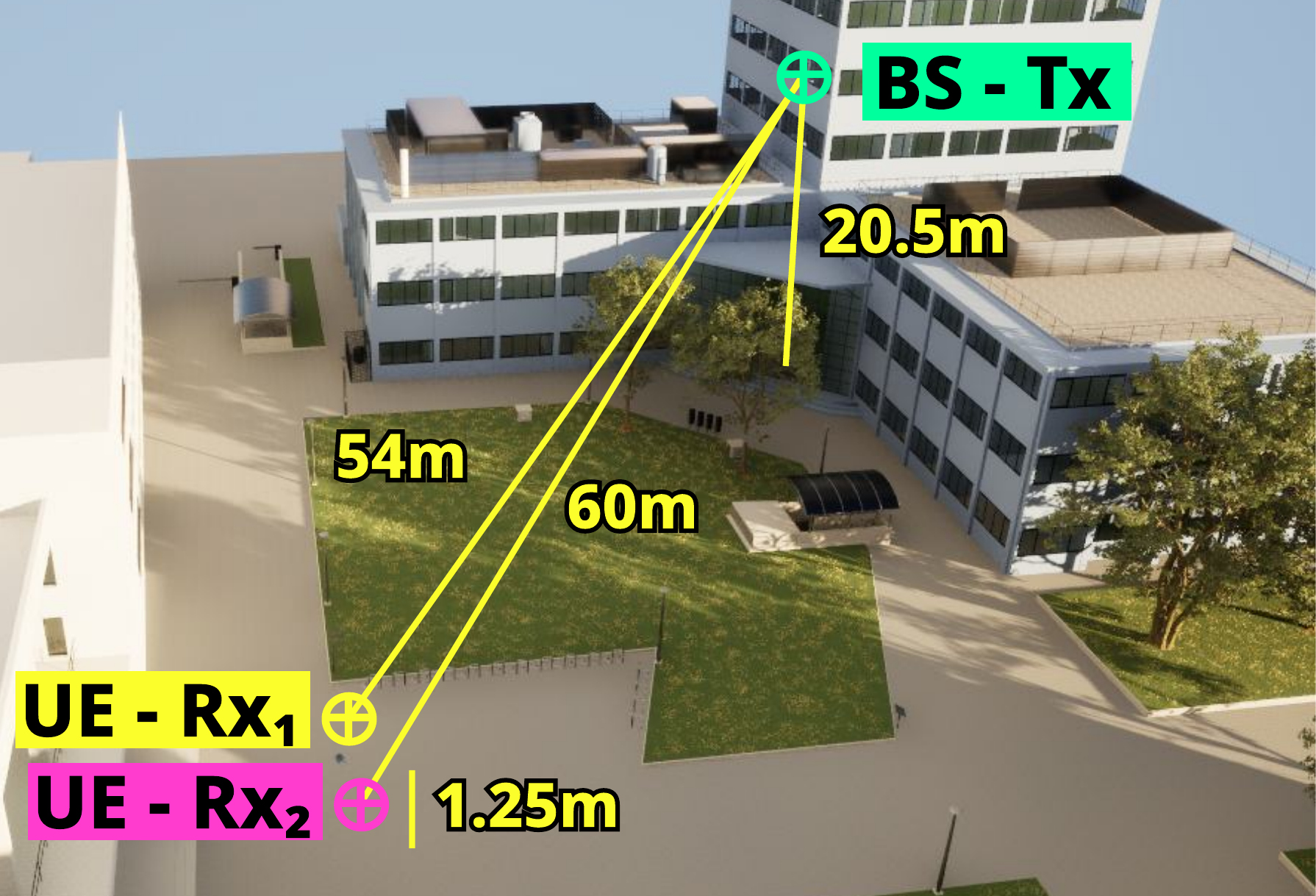}
    \label{fig:b1b2b3_pos}
  \end{subfigure}\hfill
  \begin{subfigure}[b]{0.2357\textwidth}
    \includegraphics[width=\linewidth]{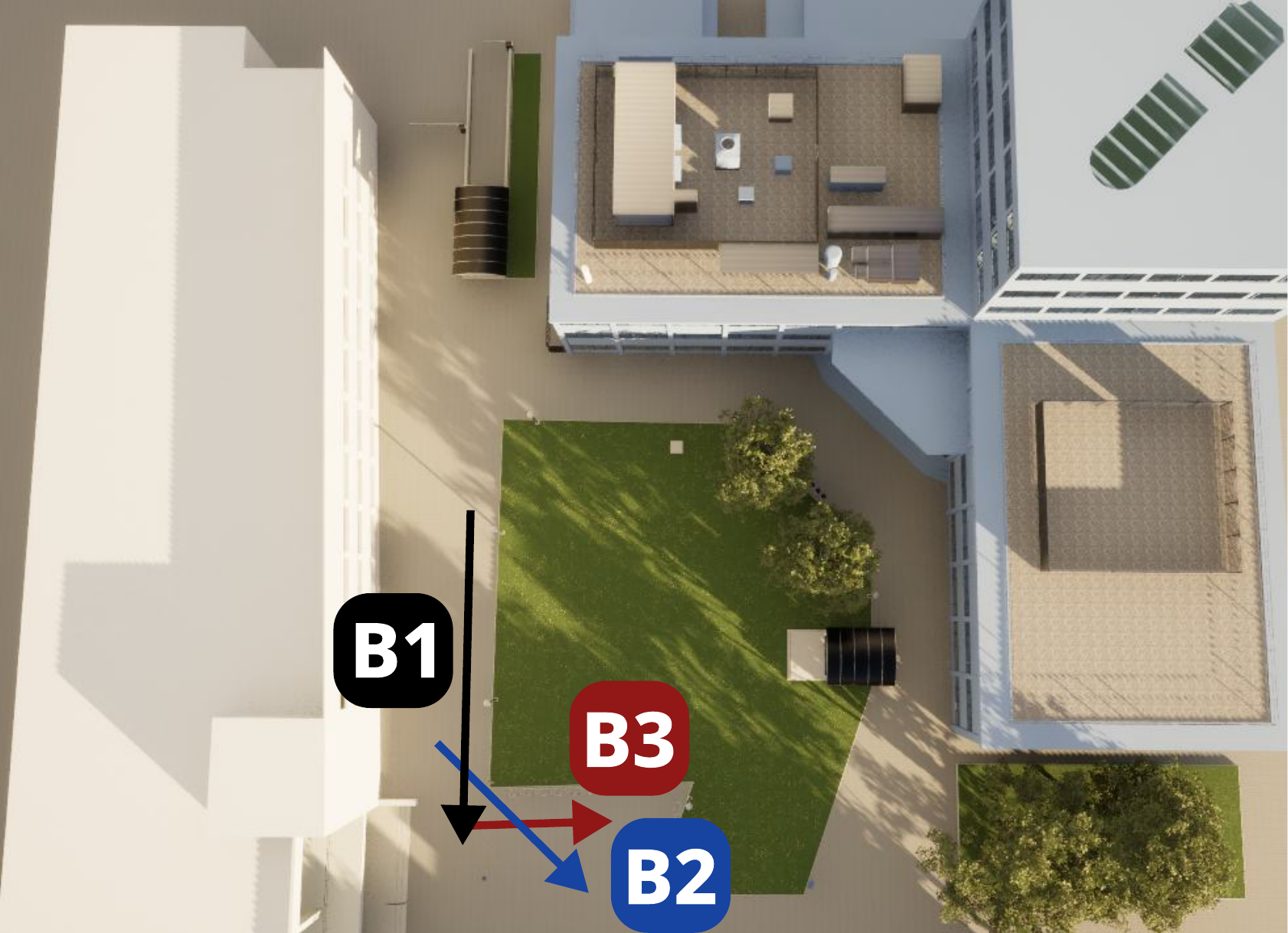}
    \label{fig:b1b2b3_traj}
  \end{subfigure}
  \begin{subfigure}[b]{0.1083\textwidth}
    \includegraphics[width=\linewidth]{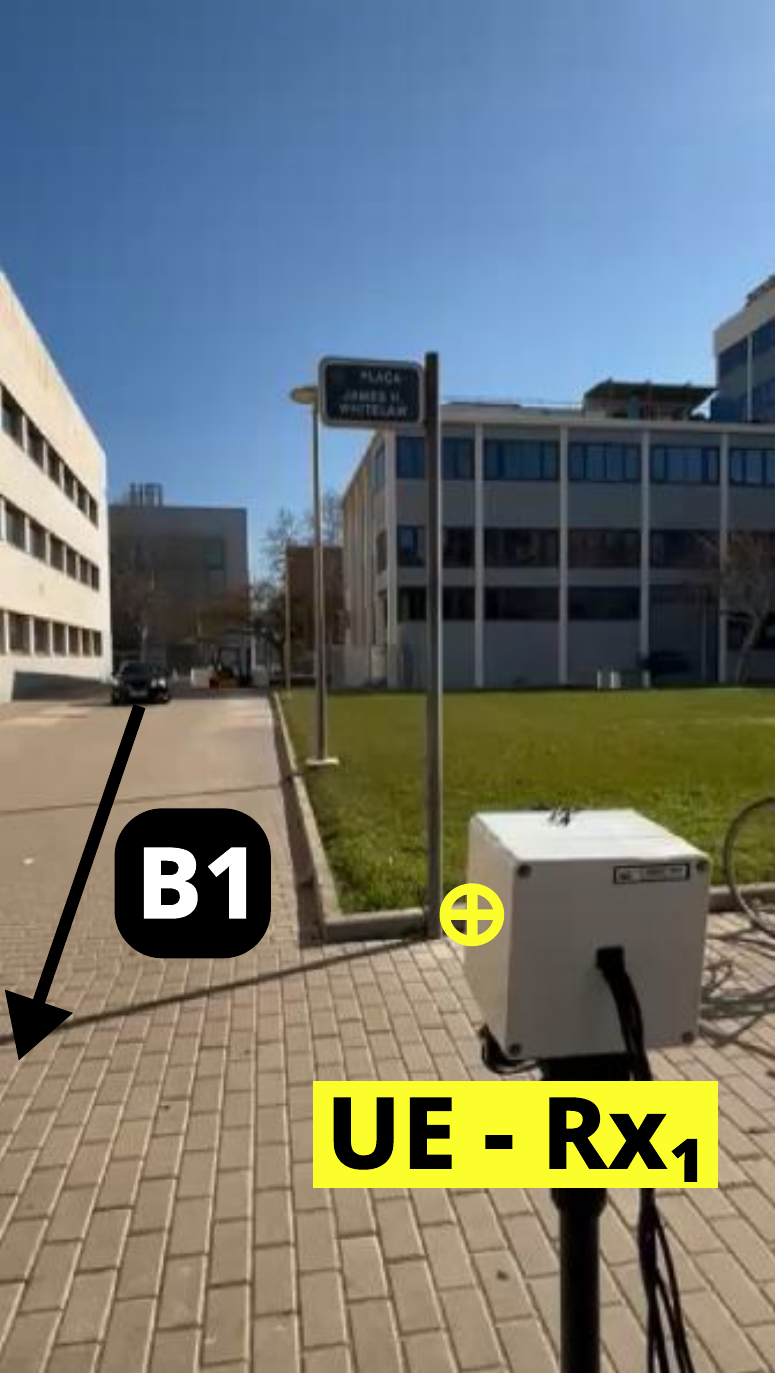}
    \label{fig:b1_photo}
  \end{subfigure}\hfill
  \begin{subfigure}[b]{0.3789\textwidth}
    \includegraphics[width=\linewidth]{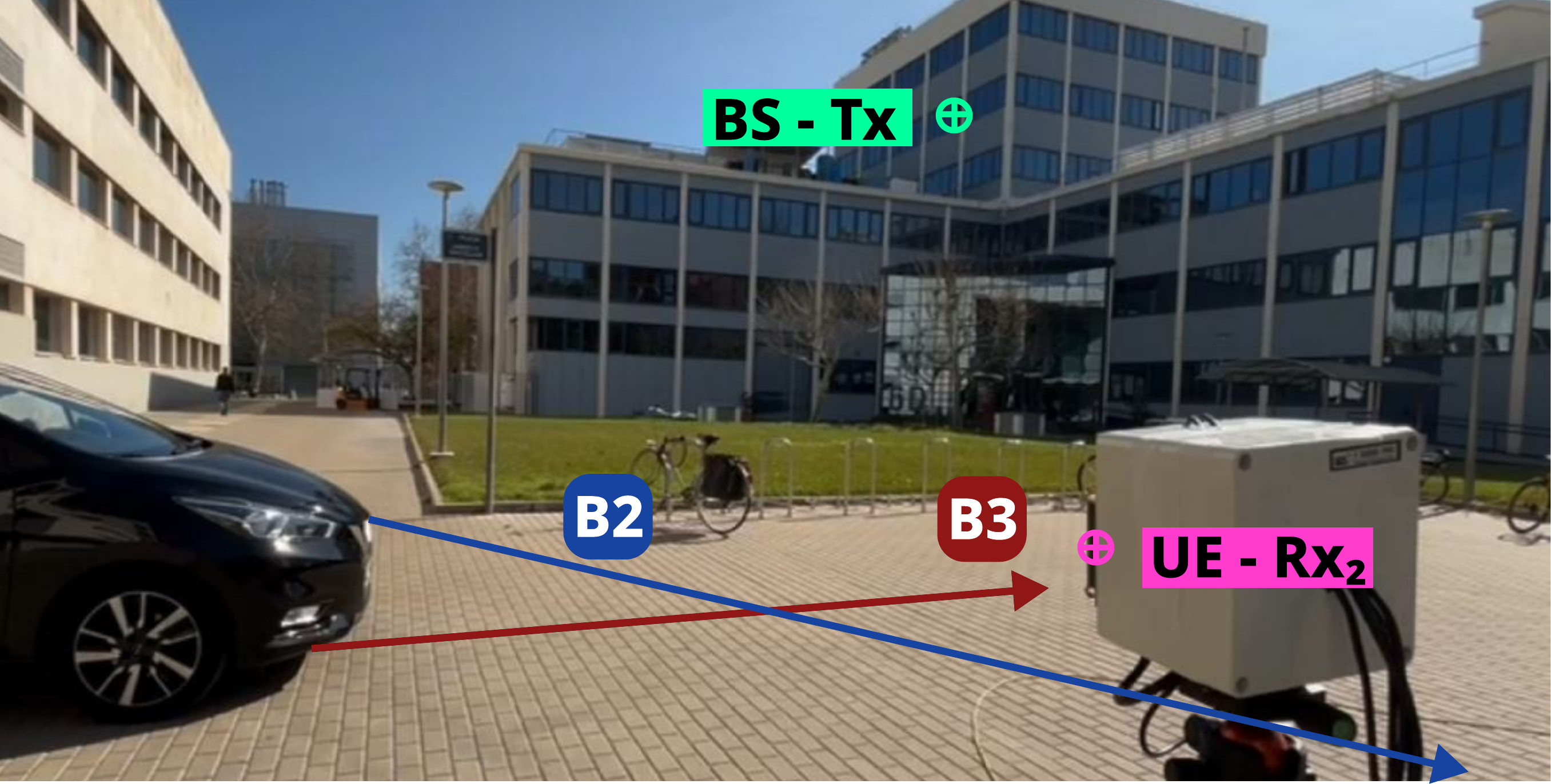}
    \label{fig:b1b2_photo}
  \end{subfigure}
  \caption{Measurement scenarios illustrating ISAC channel configuration B with Modes 2 and 5 for trajectories B1, B2 and B3.}
  \label{fig:scenariosB}
\end{figure}

Three measurements were conducted under Configuration B, in which both the transmitter and receiver remained static while the target was in motion. All measurements in this configuration were acquired from the UE side, enabling the evaluation of Mode 2, corresponding to mono-static sensing at the UE, and Mode 5, corresponding to bi-static sensing of the downlink BS-UE communication. The receiver was positioned at two distinct locations: in the first position, only trajectory B1 was recorded, where the target approached the receiver along the antenna's boresight direction. In the second position, trajectories B2 and B3 were measured, corresponding to target movements along oblique and perpendicular directions relative to the antenna's boresight, respectively. The target model used in all measurements under this configuration was a 2018 Nissan Micra. The distances and positions of every element, as well as the target trajectories, are depicted in the schematics and photographs shown in Fig.~\ref{fig:scenariosB}.


\vspace{3pt}
\subsubsection{Scenarios C}
\begin{figure}[htbp]
  \centering
  \begin{subfigure}[b]{0.2506\textwidth}
    \includegraphics[width=\linewidth]{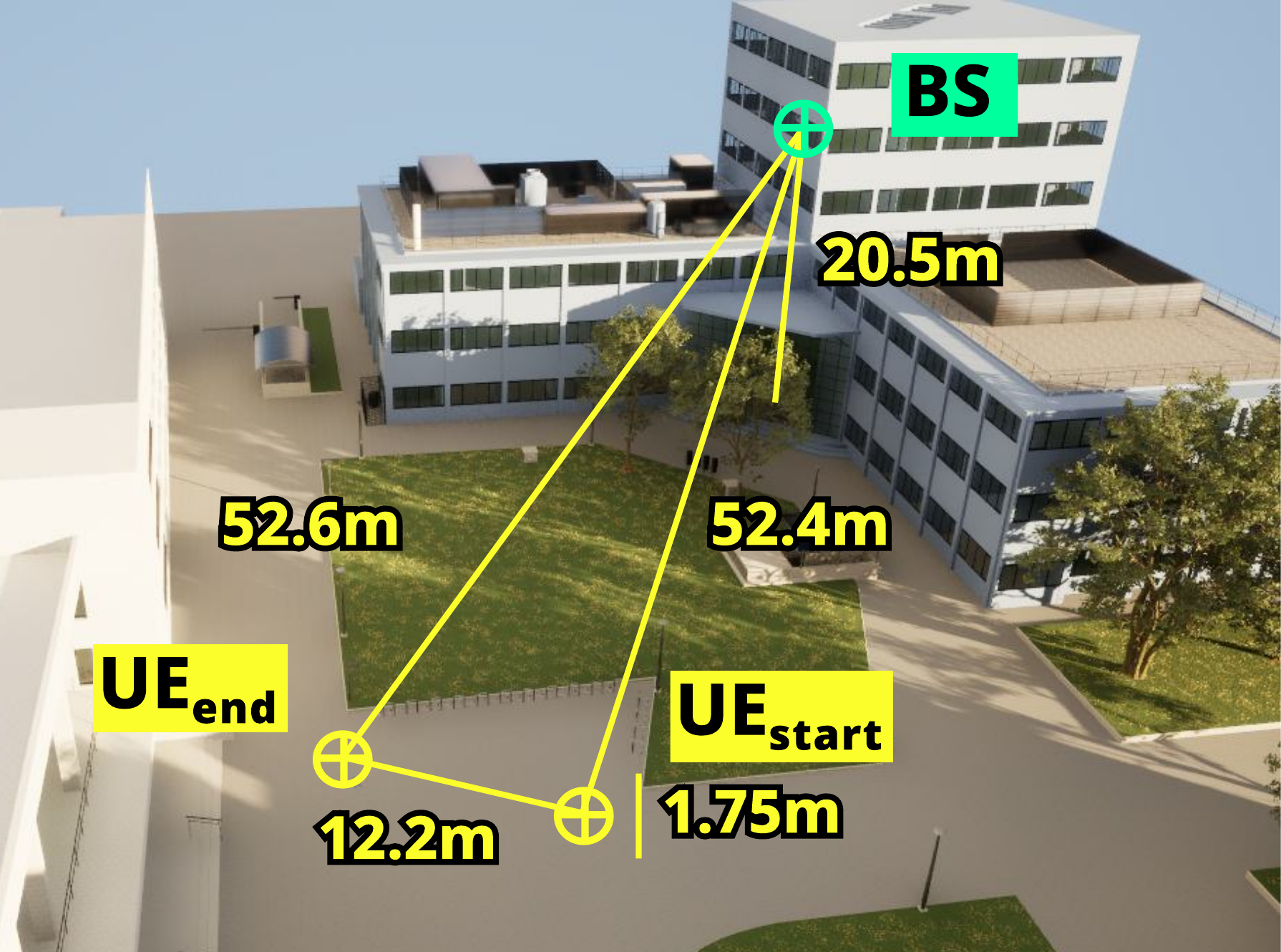}
    \label{fig:c1c3_pos}
  \end{subfigure}\hfill
  \begin{subfigure}[b]{0.2357\textwidth}
    \includegraphics[width=\linewidth]{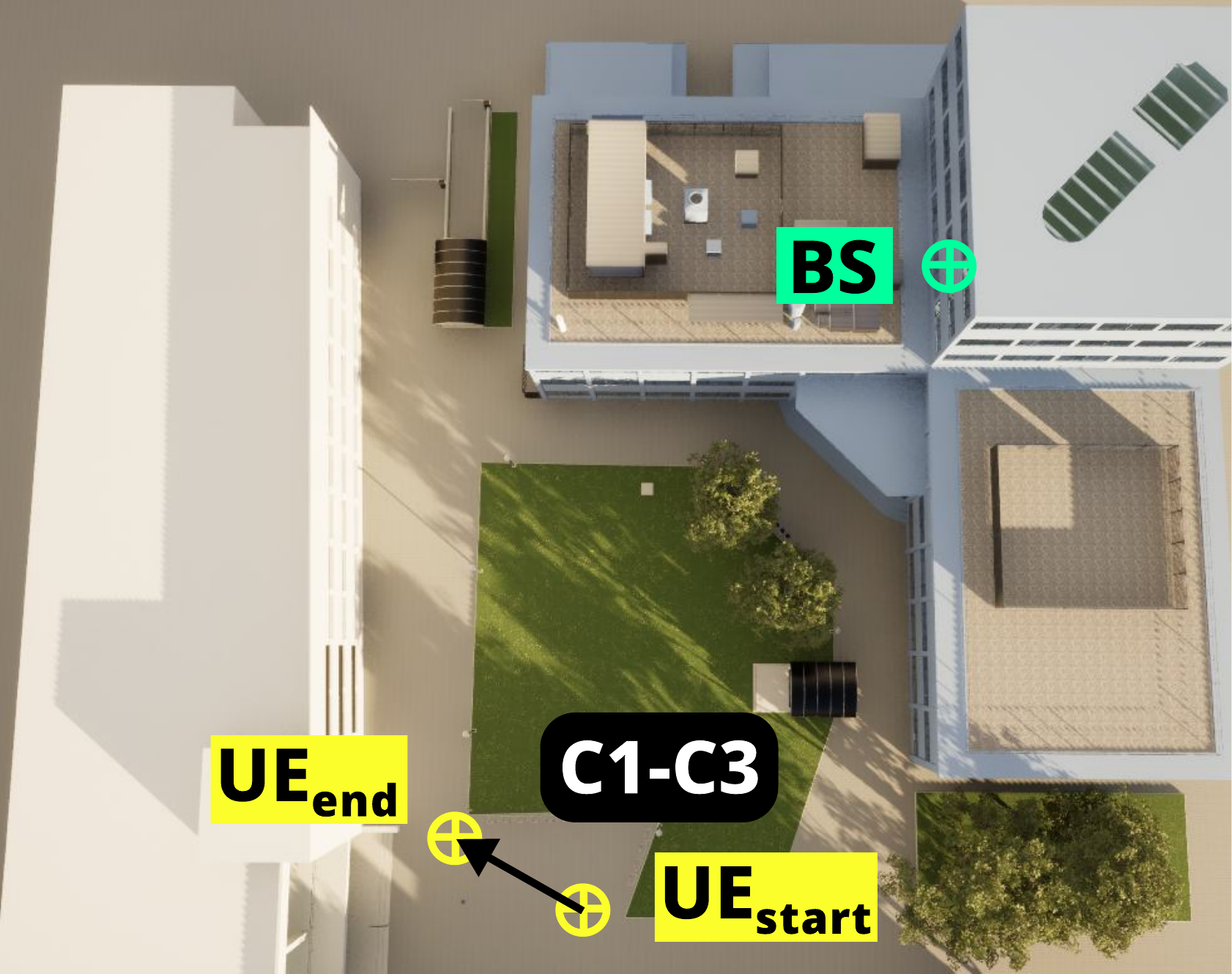}
    \label{fig:c1c3_traj}
  \end{subfigure}
  \begin{subfigure}[b]{0.2506\textwidth}
    \includegraphics[width=\linewidth]{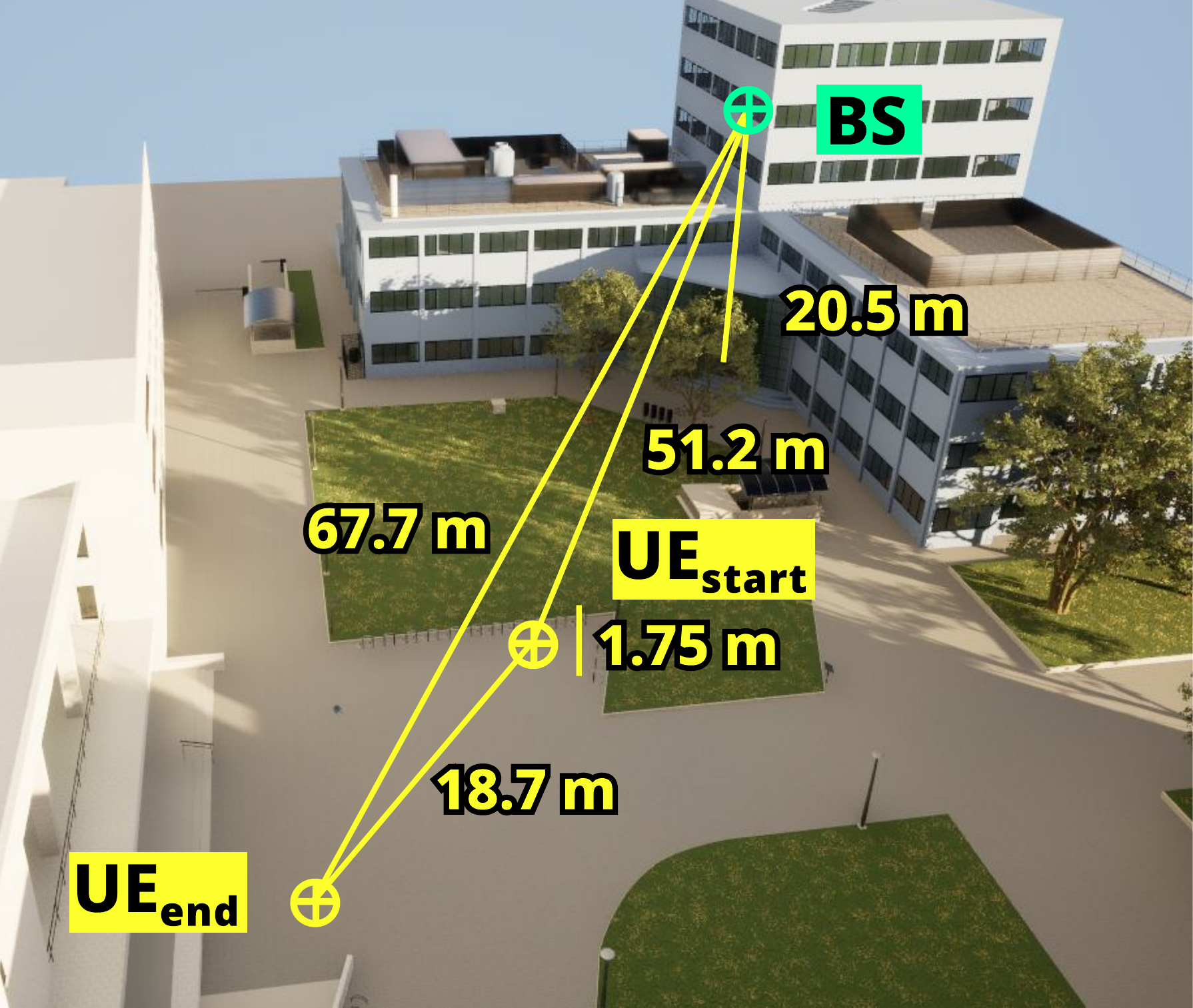}
    \label{fig:c2c4_pos}
  \end{subfigure}\hfill
  \begin{subfigure}[b]{0.2357\textwidth}
    \includegraphics[width=\linewidth]{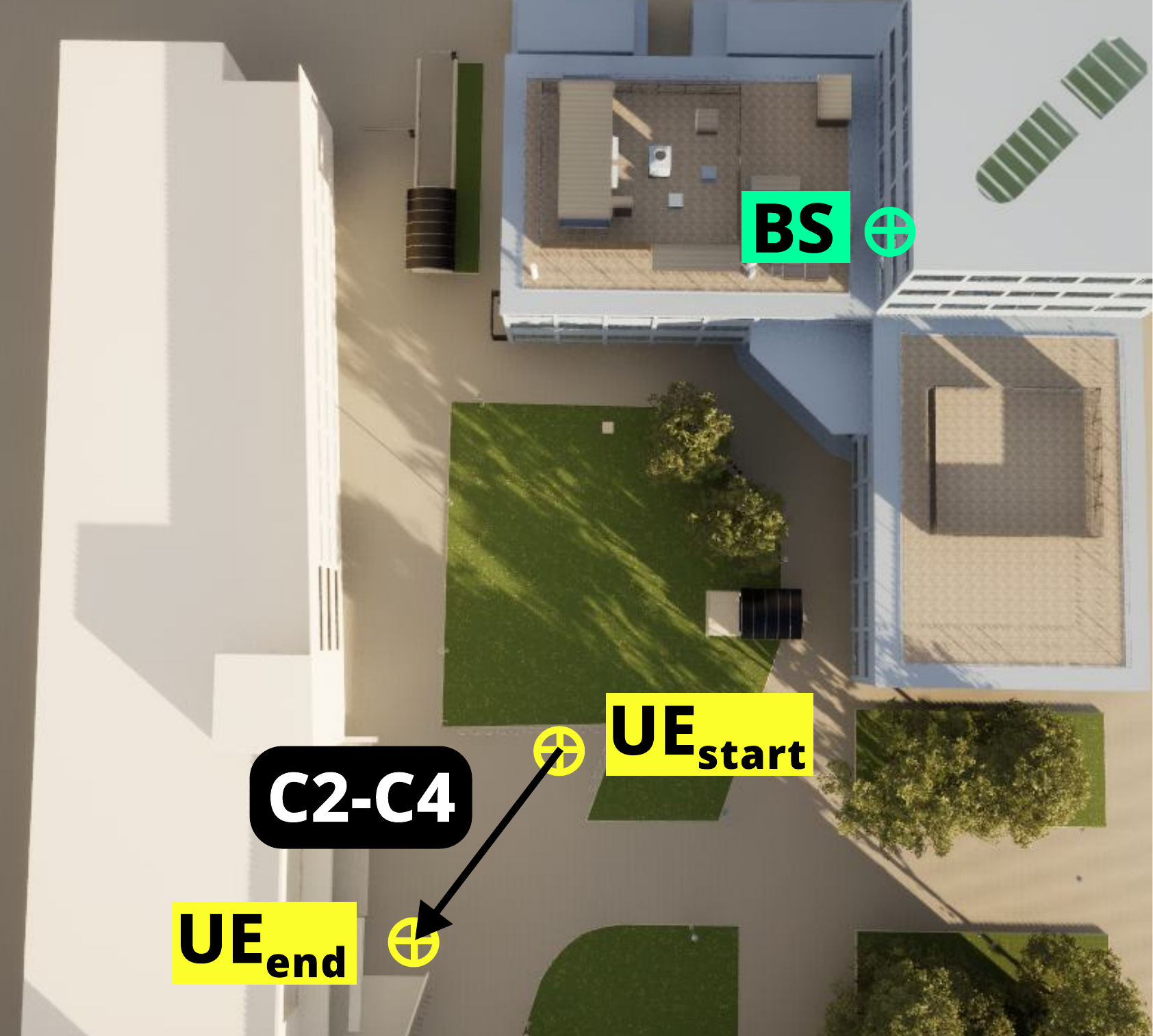}
    \label{fig:c2c4_traj}
  \end{subfigure}
  \caption{Measurement scenarios illustrating ISAC channel configuration C with Modes 1 and 5 (trajectories C1 and C3) and with Modes 2 and 5 (trajectories C2 and C4).}
  \label{fig:scenariosC}
\end{figure}

Four measurements were conducted under Configuration C, in which either the transmitter or the receiver was in motion. This measurement set includes two types of trajectories captured from both the BS side (trajectories C1 and C2) and the UE side (trajectories C3 and C4). In all cases, the UE was mounted on the roof of the moving vehicle. This approach enabled the study of Mono-static Modes 1 and 2, corresponding to mono-static sensing from the BS and UE, respectively; as well as the characterization of bi-static sensing for the downlink/uplink BS-UE communication. The two types of trajectories mentioned correspond, first, to the car moving perpendicular to the BS antenna’s boresight direction, and second, to the car moving parallel to the boresight projected direction while receding from the BS. The sensing target used in trajectories C1 and C2 was a Volvo XC60, while a KIA Xceed was employed in trajectories C3 and C4. The positions of the BS and UE, along with the specific trajectories followed by the target, are illustrated in the schematics and photographs provided in Fig.~\ref{fig:scenariosC}.


\section{Dual-Channel Analysis Metrics}
\begin{figure}[htbp]
  \centering
  \begin{subfigure}[b]{0.233\textwidth}
    \includegraphics[width=\linewidth]{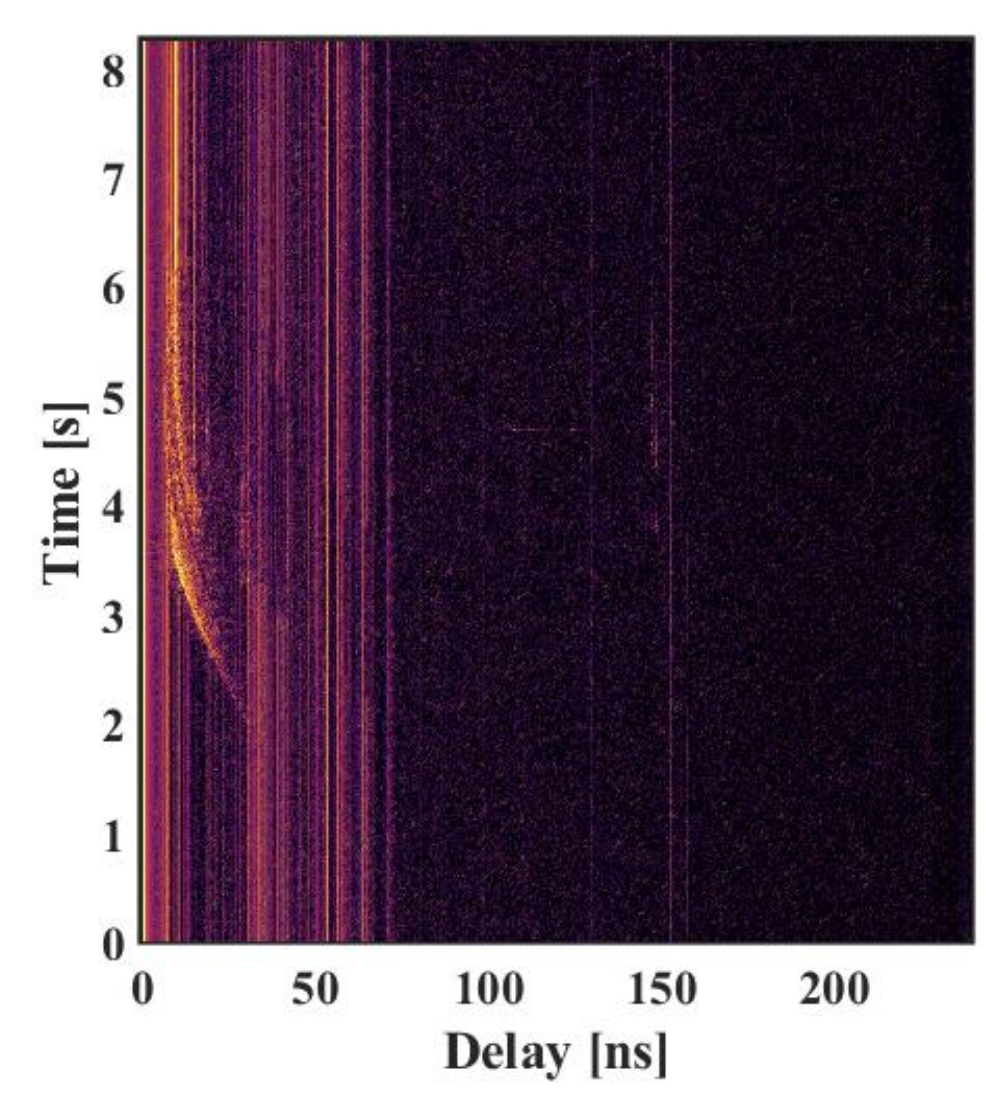}
    \caption{}
    \label{fig:mode2eg}
  \end{subfigure}\hfill
  \begin{subfigure}[b]{0.254\textwidth}
    \includegraphics[width=\linewidth]{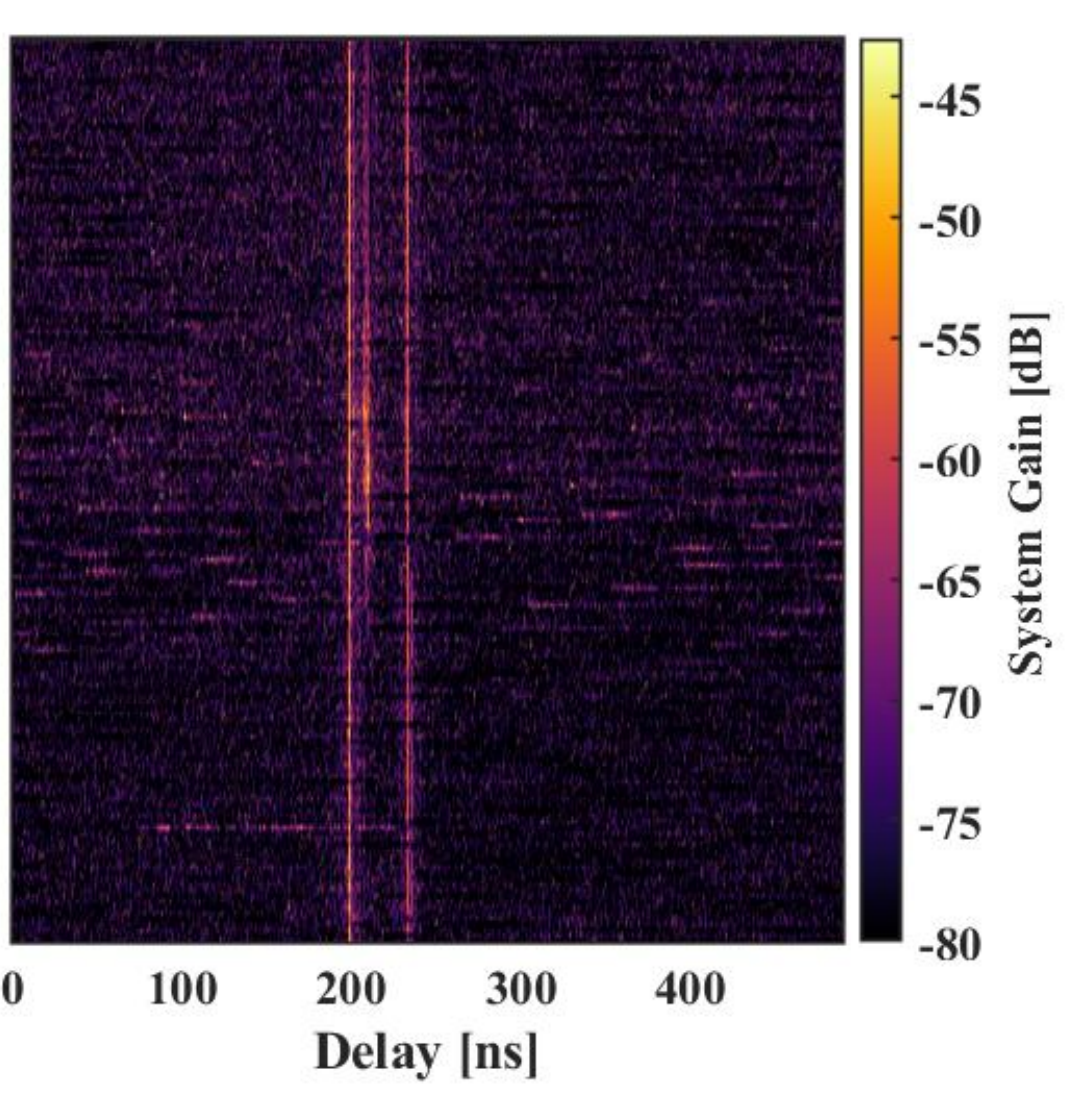}
    \caption{}
    \label{fig:mode5eg}
  \end{subfigure}
  \caption{Measured PDP for Scenario B2 in (a) Mode 2 and (b) Mode 5.}
  \label{fig:PDPeg}
\end{figure}

This section presents the metrics developed to assess the relationship between mono-static and bi-static sensing configurations within a unified UMi framework. The analysis relies on the Power Delay Profile (PDP), directly extracted from the measurements conducted using a specialized FMCW sensor-based channel sounding system. The PDP provides a time-resolved representation of the received signal power as a function of propagation delay, allowing for a detailed characterization of the multipath propagation environment. An illustrative example of the measured PDPs for both mono-static (Mode 2) and bi-static (Mode 5) sensing is shown in Fig.~\ref{fig:PDPeg}, corresponding to a representative instance from Scenario B2.

To systematically examine the interplay between mono-static and bi-static channels, we define two complementary perspectives: (i) the instantaneous correlation between mono- and bi-static channel realizations observed at a given time, and (ii) the temporal consistency of such relationships across a measurement sequence. These perspectives are formalized through the metrics introduced in the following subsections. All analyses are conducted using LOS measurements under both Configuration B and C, as described in Section~\ref{sec:measurement_scenarios}.

\subsection{Instantaneous Cross-Mode Correlation}

Instantaneous similarity between mono- and bi-static channels is quantified by the Pearson correlation coefficient evaluated over the delay domain \(\tau\in[0,T_{\max}]\). Let \(P_m(t,\tau)\) and \(P_b(t,\tau)\) denote the absolute PDP magnitudes at time \(t\). Their delay-domain means are

\[
\resizebox{\linewidth}{!}{$
\bar P_{m}(t)=\frac{1}{T_{\max}}\int_{0}^{T_{\max}} P_m(t,\tau)\,d\tau,
\qquad
\bar P_{b}(t)=\frac{1}{T_{\max}}\int_{0}^{T_{\max}} P_b(t,\tau)\,d\tau.
$}
\]

\newpage
The correlation coefficient is then

\begin{equation}
\resizebox{\linewidth}{!}{$
\rho(t)=
\frac{\displaystyle\int_{0}^{T_{\max}}\!\bigl[P_m(t,\tau)-\bar P_{m}(t)\bigr]\bigl[P_b(t,\tau)-\bar P_{b}(t)\bigr]\,d\tau}
{\sqrt{\displaystyle\int_{0}^{T_{\max}}\!\bigl[P_m(t,\tau)-\bar P_{m}(t)\bigr]^{2}d\tau}\;
 \sqrt{\displaystyle\int_{0}^{T_{\max}}\!\bigl[P_b(t,\tau)-\bar P_{b}(t)\bigr]^{2}d\tau}}.
$}
\label{eq:PearsonCont}
\end{equation}

Values of \(\rho(t)\) approaching unity indicate strong instantaneous alignment of the multipath structure, whereas lower values reflect decorrelation.

\subsection{Temporal Consistency}

Temporal consistency metrics describe how rapidly a given channel evolves \cite{Umansky2009Stationarity,Tan2017Comparison}. A symmetric sliding window of half-length \(T_{w}\) (in seconds) is centered at each instant \(t\). For any delay \(\tau\), the window-averaged magnitude is

\[
\mu(t,\tau)=\frac{1}{2T_{w}}\int_{t-T_{w}}^{t+T_{w}} P(u,\tau)\,du.
\]

\paragraph{Standard Deviation in Sliding Window}

The standard deviation within the window, averaged over the delay domain, is defined as

\begin{equation}
\resizebox{\linewidth}{!}{$
\sigma(t)=\frac{1}{T_{\max}}\int_{0}^{T_{\max}}
\sqrt{\frac{1}{2T_{w}}\int_{t-T_{w}}^{t+T_{w}}\!\!\!\bigl[P(u,\tau)-\mu(t,\tau)\bigr]^{2}du}\;d\tau.
$}
\label{eq:StdWin}
\end{equation}

Higher \(\sigma(t)\) indicates larger long-term fluctuations, making this metric particularly relevant to address main channel changes.

\paragraph{Pearson Correlation in Sliding Window}

Using the coefficient in Eq.~\eqref{eq:PearsonCont}, temporal stability is further quantified by correlating the instantaneous profile \(P(t,\tau)\) with its local mean \(\mu(t,\tau)\):

\[
\resizebox{\linewidth}{!}{$
c(t)=
\frac{\displaystyle\int_{0}^{T_{\max}}\!\bigl[P(t,\tau)-\bar P(t)\bigr]\bigl[\mu(t,\tau)-\bar\mu(t)\bigr]\,d\tau}
{\sqrt{\displaystyle\int_{0}^{T_{\max}}\!\bigl[P(t,\tau)-\bar P(t)\bigr]^{2}d\tau}\;
 \sqrt{\displaystyle\int_{0}^{T_{\max}}\!\bigl[\mu(t,\tau)-\bar\mu(t)\bigr]^{2}d\tau}},
$}
\]

where \(\bar P(t)\) and \(\bar\mu(t)\) are delay-domain means of \(P(t,\tau)\) and \(\mu(t,\tau)\). The coefficient \(c(t)\in[0,1]\) delivers a normalized measure of predictability: values near unity denote high temporal coherence, whereas smaller values reveal rapid channel variation \cite{Blumenstein2017Spatial}.

For all subsequent evaluations, the temporal window length was set to 10\% of the total acquisition time: about \(2T_w \approx 0.8\,\mathrm{s}\) for the Scenarios B and \(2T_w \approx 1.0\,\mathrm{s}\) for the Scenarios C. This choice ensures sufficient temporal resolution while enabling robust tracking of gradual channel dynamics.

\section{Results}
\subsection{Instantaneous Cross-Mode Correlation Analysis}
Figure~\ref{fig:PDPeg} provides the PDPs for Scenario B2, comparing mono-static sensing (Mode 2) and bi-static sensing (Mode 5). A visual inspection reveals considerable structural divergence in the multipath components across the two modes. The dominant paths and delay spreads differ markedly, consistent with their distinct propagation geometries-round-trip versus forward-reverse trajectories.

\begin{figure}[htbp]
  \centering
  \begin{subfigure}[b]{0.48\textwidth}
    \includegraphics[width=\linewidth]{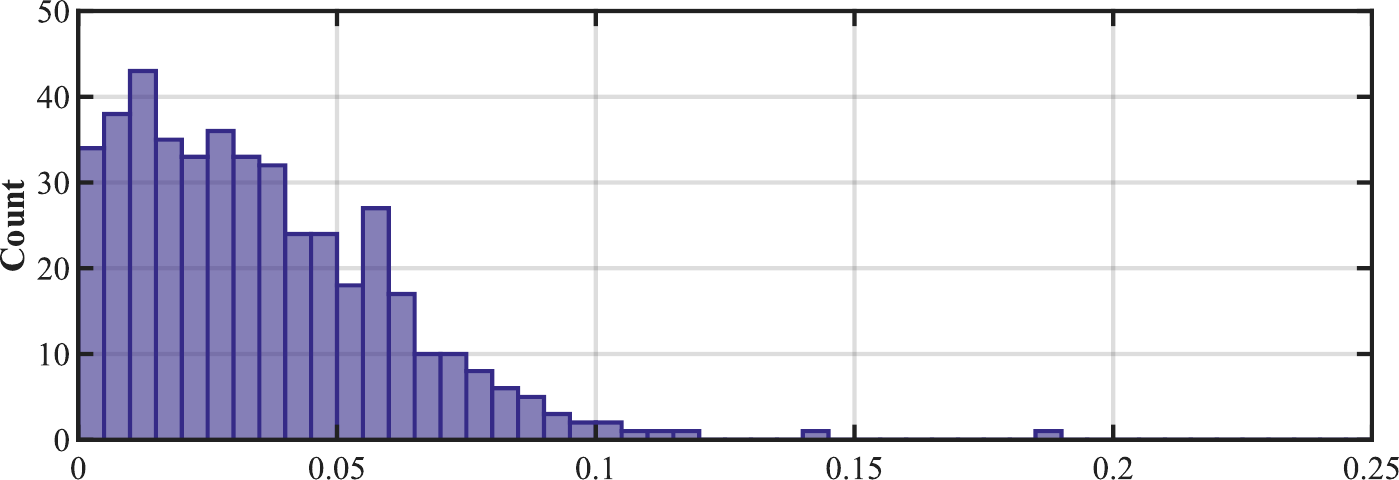}
    \caption{}
    \label{fig:corrB}
  \end{subfigure}\vfill
  \begin{subfigure}[b]{0.48\textwidth}
    \includegraphics[width=\linewidth]{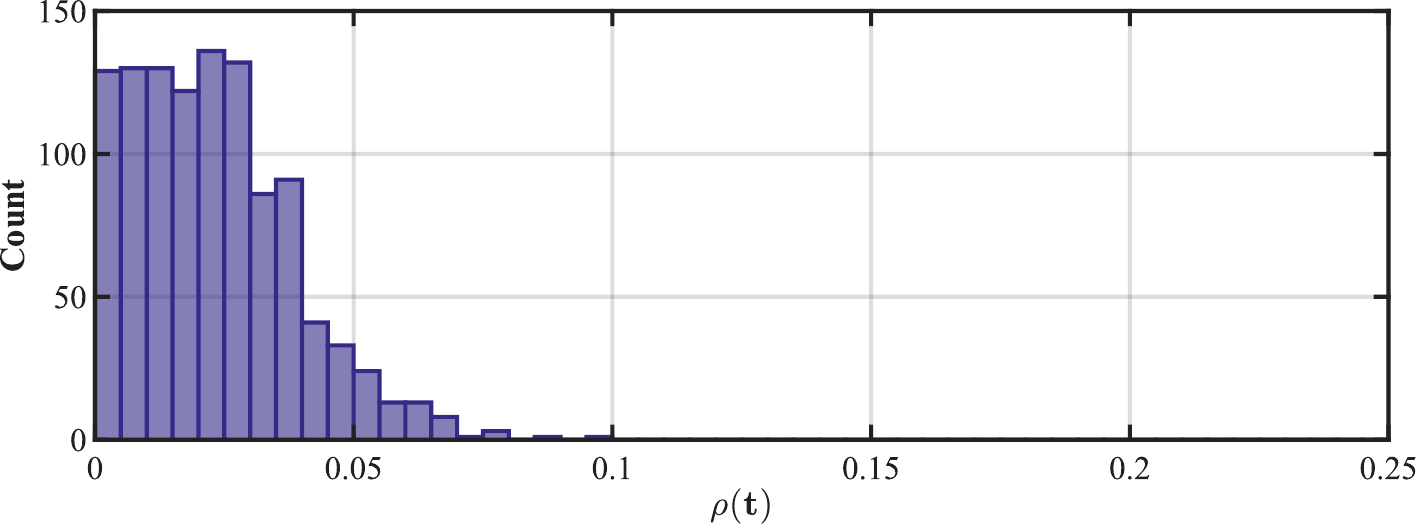}
    \caption{}
    \label{fig:corrC}
  \end{subfigure}
  \caption{Histograms of the instantaneous Cross-Mode correlation between Mono‐static and Bi‐static Modes, aggregated across (a) Scenarios B and (b) Scenarios C.}
  \label{fig:corrHist}
\end{figure}

Quantitatively, the Pearson correlation coefficient \(\rho(t)\), computed over the delay domain, remains consistently low across all measurement instants. As illustrated by the aggregated histograms in Figure~\ref{fig:corrHist}, the majority of correlation values for both Scenarios B (Figure~\ref{fig:corrB}) and Scenarios C (Figure~\ref{fig:corrC}) are concentrated between 0 and 0.05, with only infrequent instances exceeding 0.1 and rare peaks up to 0.2. Visually, the correlation values for Scenarios B appear marginally higher than those for Scenarios C, though this difference is not substantial enough to warrant a definitive conclusion. These results support the first major finding: mono-static and bi-static channels exhibit low instantaneous correlation, even under LOS conditions. This confirms that direct substitution or extrapolation between channel types is infeasible for real-time ISAC operations, as instantaneous realizations are not reliably interchangeable.

\subsection{Temporal Consistency Across Configurations}
\begin{figure}[htbp]
  \centering
  \begin{subfigure}[b]{0.48\textwidth}
    \includegraphics[width=\linewidth]{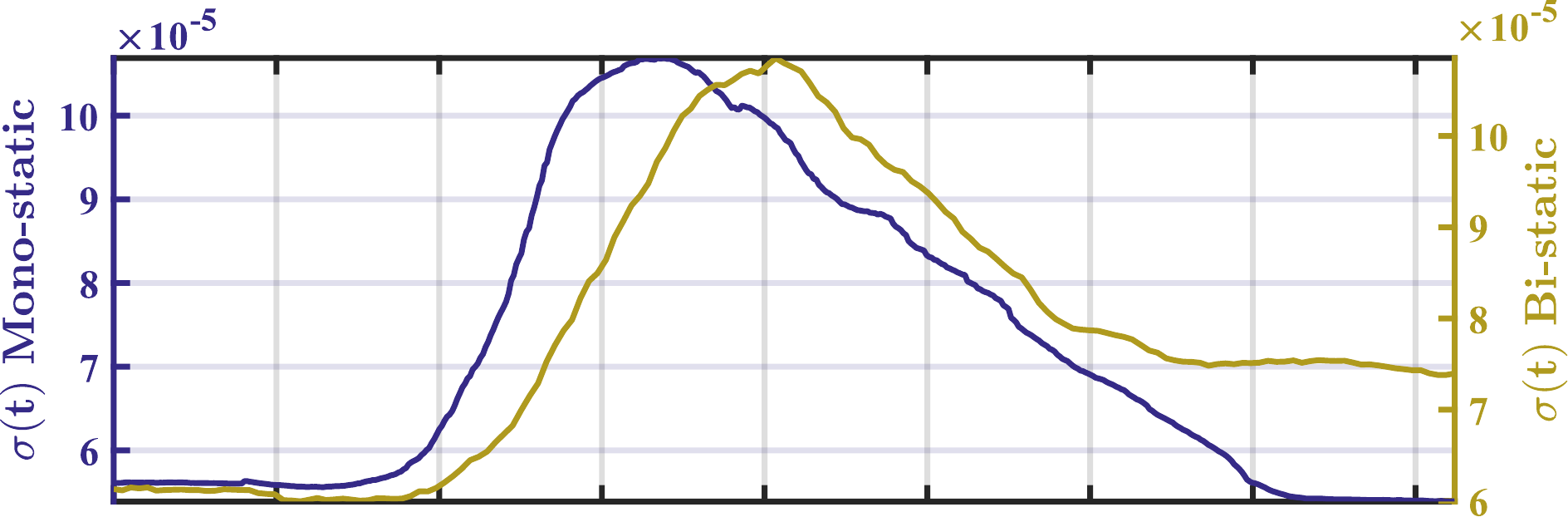}
    \caption{}
    \label{fig:stdB2}
  \end{subfigure}\vfill
  \begin{subfigure}[b]{0.48\textwidth}
    \includegraphics[width=\linewidth]{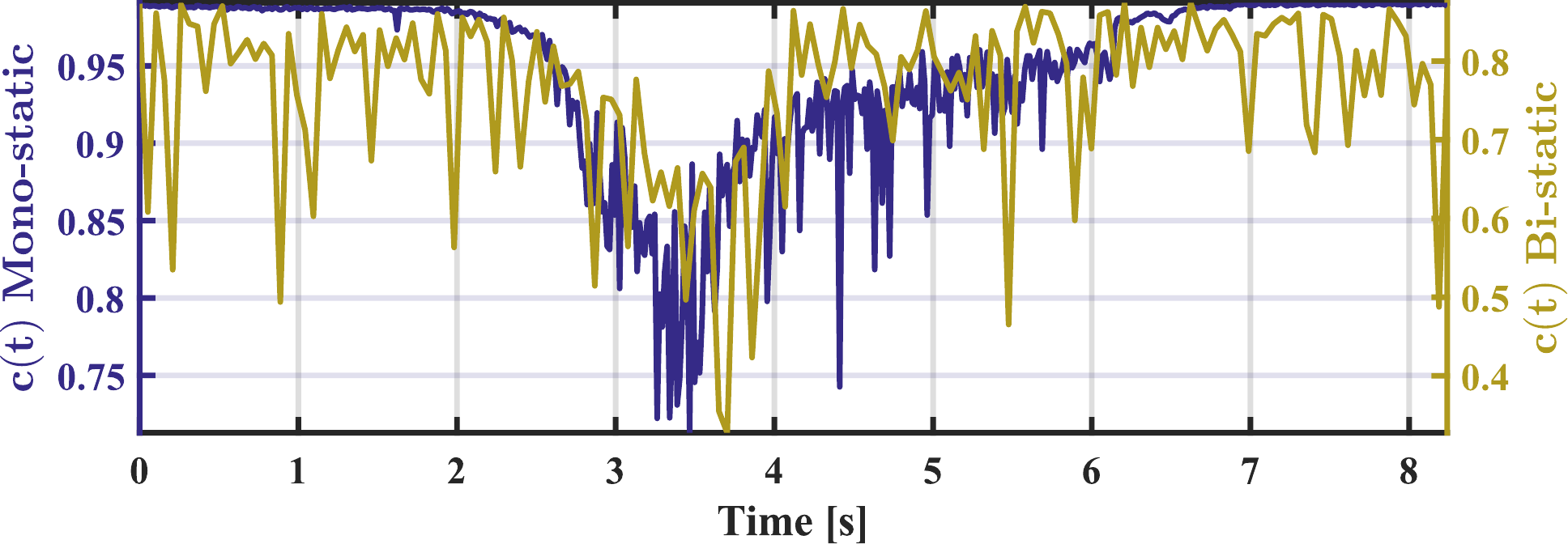}
    \caption{}
    \label{fig:pearsonB2}
  \end{subfigure}
  \caption{Temporal consistency metrics. (a) Standard deviation and (b) Pearson correlation, in sliding window over the time domain.}
  \label{fig:consB2}
\end{figure}

Building again upon the PDPs for Scenario B2 depicted in Figure~\ref{fig:PDPeg}, the analysis is extended to temporal consistency. Despite the low instantaneous correlation, Figure~\ref{fig:consB2} highlights a remarkable level of temporal consistency in the evolution of both channel types.

Figure~\ref{fig:stdB2} plots the delay-averaged standard deviation \(\sigma(t)\) over a symmetric sliding window, revealing smooth temporal variations for both mono-static and bi-static channels. Peaks in \(\sigma(t)\) align with moments of rapid environmental change (e.g., target proximity), but overall fluctuations remain bounded. This suggests the channel dynamics are primarily governed by macro-level kinematics, such as the vehicle's motion trajectory, rather than fine-grained environmental noise.

Figure~\ref{fig:pearsonB2} illustrates the Pearson correlation coefficient \(c(t)\) between the instantaneous PDP and its local temporal average. For both sensing modes, \(c(t)\) consistently exceeds 0.8, particularly in segments of steady motion. This confirms that despite structural decorrelation, both mono-static and bi-static channels evolve in a temporally coherent and predictable manner. This predictability holds across the three trajectories measured under B configurations, for which the mono-static sensing was captured from the UE (Mode 2).

\begin{figure}[htbp]
  \centering
  \begin{subfigure}[b]{0.249\textwidth}
    \includegraphics[width=\linewidth]{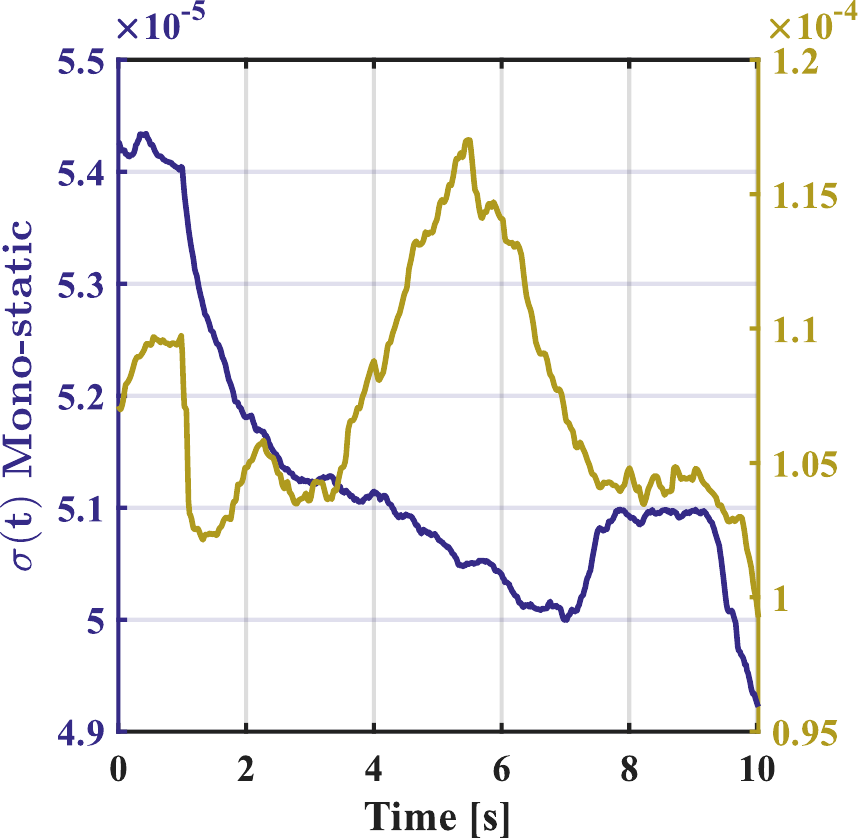}
    \caption{}
    \label{fig:c2cons}
  \end{subfigure}\hfill
  \begin{subfigure}[b]{0.238\textwidth}
    \includegraphics[width=\linewidth]{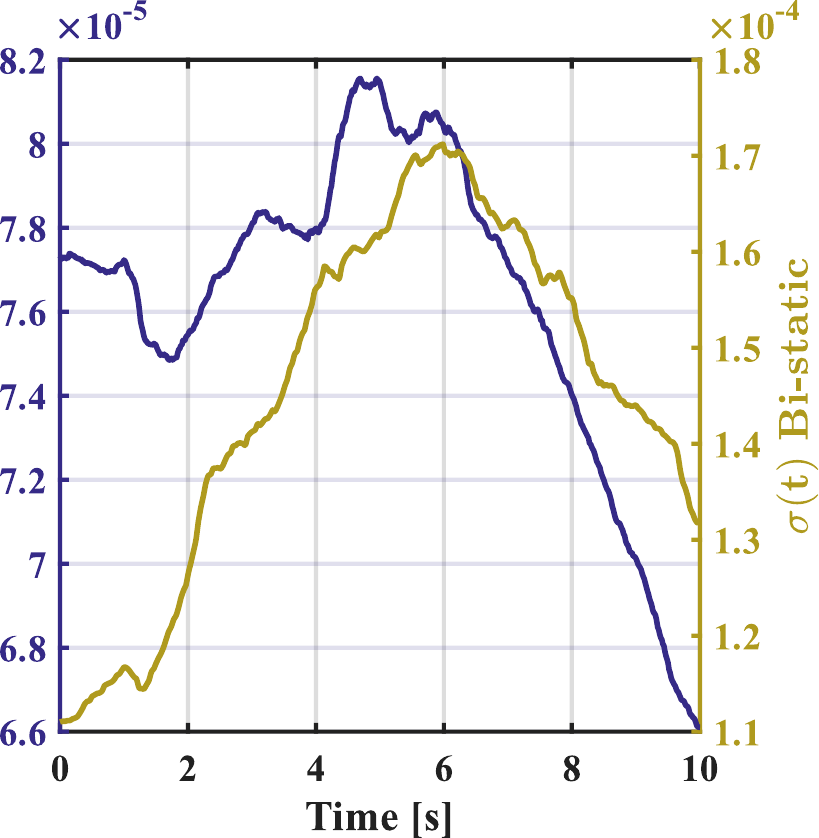}
    \caption{}
    \label{fig:c4cons}
  \end{subfigure}
  \caption{Standard deviation consistency metric for (a) Scenario C2 and (b) Scenario C4, in sliding window over the time domain. }
  \label{fig:ccons}
\end{figure}

Further characterization of temporal consistency is achieved by examining \(\sigma(t)\) for Scenarios C2 and C4 (Figure~\ref{fig:ccons}). Both scenarios share an identical target trajectory; however, Scenario C2 captures mono-static sensing from the BS side (Mode 1), whereas Scenario C4 captures mono-static sensing from the UE side (Mode 2). In both cases, the bi‑static channels display remarkably similar temporal trends, consistent with the theoretical reciprocity of bi‑static propagation. Moreover, the mono‑static channel in Scenario C4 (Mode 2) closely follows these bi‑static trends, indicating highly predictable behavior across both channel types. In contrast, the mono‑static channel in Scenario C2 (Mode 1) deviates from this pattern, exhibiting temporal dynamics that do not align with the bi‑static evolution. These observations suggests that for Modes 2 and 5, temporal variations are strongly predictable and this conclusion appears generalizable across different trajectories and configurations. However, for Modes 1 and 5, a clear and strong relationship remains less evident.

\section{Conclusions}

A comprehensive examination of the seven UMi measurement trajectories, covering both static-Tx/Rx with dynamic targets (Configuration B) and dynamic-Tx or Rx setups (Configuration C), confirms two robust properties of ISAC propagation:

\begin{itemize}
\item \textbf{Low Instantaneous Cross-Mode Correlation.}
Across every trajectory, including perpendicular and oblique motion paths, the correlation between mono-static and bi-static PDPs remains tightly clustered below 0.1. This finding is consistent for Mono-static Modes 1 and 2 and for Bi-static Mode 5, demonstrating that the divergent round-trip versus forward-reverse geometries lead to fundamentally decorrelated instantaneous multipath structures.

\item \textbf{Conditional Unified Temporal Consistency.}
When temporal windows are applied, a clear dichotomy emerges. For the five measurements that involve Mono-static Mode 2 and Bi-static Mode 5 (trajectories B1--B3, C3, and C4), the sliding-window standard deviation $\sigma(t)$ evolves in lock-step that evidences a shared, predictable temporal evolution that can be exploited for channel tracking and waveform adaptation.  
Conversely, Scenarios C1 and C2, which employ Mono‑static Mode 1 at the BS, depart from this coherent trend. In these two runs the vehicle reflections contribute markedly less power than static surroundings (walls, ground, building facades). As a result, the mono‑static channel is dominated by quasi‑static components whose slow variations decouple from the bi‑static link that remains sensitive to the forward‑reverse path to the moving target. 
\end{itemize}

The implications for 6G ISAC design are threefold. (1) The low instantaneous correlation between mono-static and bi-static sensing dictates that waveforms and algorithms must be mode-specific; solutions developed for one mode cannot be transferred blindly to another. (2) The unified temporal evolution observed in Modes 2 and 5 opens the door to predictive channel-tracking schemes that extrapolate future states from only a brief measurement history, thereby reducing sensing overhead. (3) The loss of temporal coherence in Mode 1 when the target return is weak, as evidenced in Scenarios C1 and C2, highlights the need for refined propagation models that explicitly capture situations in which static background scatterers dominate the mono-static response.

Overall, the study closes a critical empirical gap by demonstrating that, while mono-static and bi-static channels are instantaneously decorrelated, they can, but do not always, share a common temporal rhythm. Recognizing when this rhythm holds is essential for the joint design and standardization of dual-mode ISAC systems in dynamic urban microcells.

\bibliographystyle{ieeetr}   
\bibliography{references}    

@ARTICLE{Lu2024ISAC,
  author={Lu, Shihang and Liu, Fan and Li, Yunxin and Zhang, Kecheng and Huang, Hongjia and Zou, Jiaqi and Li, Xinyu and Dong, Yuxiang and Dong, Fuwang and Zhu, Jia and Xiong, Yifeng and Yuan, Weijie and Cui, Yuanhao and Hanzo, Lajos},
  journal={IEEE Internet of Things Journal}, 
  title={Integrated Sensing and Communications: Recent Advances and Ten Open Challenges}, 
  year={2024},
  volume={11},
  number={11},
  pages={19094-19120},
  doi={10.1109/JIOT.2024.3361173}}

@misc{zhang2025integratedsensingcommunicationsyears,
      title={Integrated Sensing and Communications Over the Years: An Evolution Perspective}, 
      author={Di Zhang and Yuanhao Cui and Xiaowen Cao and Nanchi Su and Fan Liu and Xiaojun Jing and J. Andrew Zhang and Jie Xu and Christos Masouros and Dusit Niyato and Marco Di Renzo},
      year={2025},
      eprint={2504.06830},
      archivePrefix={arXiv},
      primaryClass={eess.SP},
      url={https://arxiv.org/abs/2504.06830}, 
}

@misc{meng2024integratedsensingcommunicationmeets,
      title={Integrated Sensing and Communication Meets Smart Propagation Engineering: Opportunities and Challenges}, 
      author={Kaitao Meng and Christos Masouros and Kai-Kit Wong and Athina P. Petropulu and Lajos Hanzo},
      year={2024},
      eprint={2402.18683},
      archivePrefix={arXiv},
      primaryClass={cs.IT},
      url={https://arxiv.org/abs/2402.18683}, 
}

@misc{wei2024integratedsensingcommunicationchannel,
      title={Integrated Sensing and Communication Channel Modeling: A Survey}, 
      author={Zhiqing Wei and Jinzhu Jia and Yangyang Niu and Lin Wang and Huici Wu and Heng Yang and Zhiyong Feng},
      year={2024},
      eprint={2404.17462},
      archivePrefix={arXiv},
      primaryClass={cs.NI},
      url={https://arxiv.org/abs/2404.17462}, 
}

@misc{heggo2025isacchannelmodelling,
      title={ISAC Channel Modelling -- Perspectives from ETSI}, 
      author={Mohammad Heggo and Arman Shojaeifard and Alain Mourad and Chuangxin Jiang and Ruiqi Liu and Junchen Liu},
      year={2025},
      eprint={2505.10275},
      archivePrefix={arXiv},
      primaryClass={eess.SP},
      url={https://arxiv.org/abs/2505.10275}, 
}

@misc{ge2023integratedmonostaticbistaticmmwave,
      title={Integrated Monostatic and Bistatic mmWave Sensing}, 
      author={Yu Ge and Hyowon Kim and Lennart Svensson and Henk Wymeersch and Sumei Sun},
      year={2023},
      eprint={2308.13729},
      archivePrefix={arXiv},
      primaryClass={eess.SP},
      url={https://arxiv.org/abs/2308.13729}, 
}

@misc{zhang2025jointbistaticpositioningmonostatic,
      title={Joint Bistatic Positioning and Monostatic Sensing: Optimized Beamforming and Performance Tradeoff}, 
      author={Yuchen Zhang and Hui Chen and Pinjun Zheng and Boyu Ning and Hong Niu and Henk Wymeersch and Tareq Y. Al-Naffouri},
      year={2025},
      eprint={2503.03220},
      archivePrefix={arXiv},
      primaryClass={eess.SP},
      url={https://arxiv.org/abs/2503.03220}, 
}

@misc{wang2025sensingassistedchannelestimationofdm,
      title={Sensing-Assisted Channel Estimation for OFDM ISAC Systems: Framework, Algorithm, and Analysis}, 
      author={Shuhan Wang and Aimin Tang and Xudong Wang and Wenze Qu},
      year={2025},
      eprint={2502.16436},
      archivePrefix={arXiv},
      primaryClass={eess.SP},
      url={https://arxiv.org/abs/2502.16436}, 
}

@misc{keskin2025bridginggapdataaidedsensing,
      title={Bridging the Gap via Data-Aided Sensing: Can Bistatic ISAC Converge to Genie Performance?}, 
      author={Musa Furkan Keskin and Silvia Mura and Marouan Mizmizi and Dario Tagliaferri and Henk Wymeersch},
      year={2025},
      eprint={2505.01280},
      archivePrefix={arXiv},
      primaryClass={eess.SP},
      url={https://arxiv.org/abs/2505.01280}, 
}

@INPROCEEDINGS{Castilla2024DualISAC,
  author    = {Alejandro Castilla and Saúl Fenollosa and Monika Drozdowska and Alejandro Lopez-Escudero and Sergio Micó-Rosa and Narcís Cardona},
  booktitle = {2024 IEEE 35th International Symposium on Personal, Indoor and Mobile Radio Communications (PIMRC)},
  title     = {Novel Approach to Dual-Channel Estimation in Integrated Sensing and Communications for 6G},
  year      = {2024},
  pages     = {1--6},
  doi       = {10.1109/PIMRC59610.2024.10817252},
}

@INPROCEEDINGS{Cardona2023Radar,
  author={Cardona, Narcis and Romero, J. Samuel and Yang, Wenfei and Li, Jian},
  booktitle={ICASSP 2023 - 2023 IEEE International Conference on Acoustics, Speech and Signal Processing (ICASSP)}, 
  title={Integrating the Sensing and Radio Communications Channel Modelling From Radar Mutual Interference}, 
  year={2023},
  volume={},
  number={},
  pages={1-5},
  doi={10.1109/ICASSP49357.2023.10095846}}

@INPROCEEDINGS{Wenfei2023ChanMeas,
  author={Yang, Wenfei and Chen, Yi and Zhang, Yunhao and Yu, Ziming and Zhang, Min},
  booktitle={2023 IEEE Globecom Workshops (GC Wkshps)}, 
  title={ISAC Channel Measurements and Modeling Methodology}, 
  year={2023},
  volume={},
  number={},
  pages={1177-1182},
  doi={10.1109/GCWkshps58843.2023.10464999}}

@ARTICLE{Blumenstein2017Spatial,
  author={Blumenstein, Jiri and Prokes, Ales and Chandra, Aniruddha and Mikulasek, Tomas and Marsalek, Roman and Zemen, Thomas and Mecklenbräuker, Christoph},
  journal={IEEE Transactions on Vehicular Technology}, 
  title={In-Vehicle Channel Measurement, Characterization, and Spatial Consistency Comparison of $\text{30}\hbox{--}\text{11 GHz}$ and $\text{55}\hbox{--}\text{65 GHz}$ Frequency Bands}, 
  year={2017},
  volume={66},
  number={5},
  pages={3526-3537},
  doi={10.1109/TVT.2016.2600101}}

@INPROCEEDINGS{Umansky2009Stationarity,
  author={Umansky, Dmitry and Patzold, Matthias},
  booktitle={GLOBECOM 2009 - 2009 IEEE Global Telecommunications Conference}, 
  title={Stationarity Test for Wireless Communication Channels}, 
  year={2009},
  volume={},
  number={},
  pages={1-6},
  doi={10.1109/GLOCOM.2009.5425841}}

@INPROCEEDINGS{Tan2017Comparison,
  author={Tan, Yi and Wang, Cheng-Xiang and Nielsen, Jesper Odum and Pedersen, Gert F.},
  booktitle={2017 13th International Wireless Communications and Mobile Computing Conference (IWCMC)}, 
  title={Comparison of stationarity regions for wireless channels from 2 GHz to 30 GHz}, 
  year={2017},
  volume={},
  number={},
  pages={647-652},
  doi={10.1109/IWCMC.2017.7986361}}

\end{document}